# Application of Selenium-82 for Short Base Neutrino Oscillations Searches

Sergei Semenov[1, *]

[1]*National Research Centre Kurchatov Institute, 123182 Kurchatov Square, 1, Moscow, Russia*

Characteristics of neutrino absorption by $^{82}$Se nucleus reaction - low threshold value, high absorption cross section of neutrinos, emitted by artificial sources, make selenium-82 perspective object to search for new neutrino species in calibration experiments. One of the interesting possibilities is usage of scintillating crystals ×. In (3+1) model for spherical geometry the expression for neutrino path length in a setup in the presence of oscillations is obtained and the scheme of experiment is proposed.

## I. INTRODUCTION

Radiochemical method of neutrino registration [1], used in calibration experiments with $^{37}$Ar, $^{51}$Cr artificial sources of neutrino, led SAGE [2] и GALLEX [3] collaborations to conclusion that about 20% deficit of counting rate in reaction of neutrino absorption by $^{71}$Ga nuclei is the case. These results are confirmed by BEST experiment [4,5]. This set of experimental data is interpreted as the occurrence of oscillations into the forth neutrino eigenstate with mass ~1 eV [6].

The BEST project installation, located in Baksan Neutrino Observatory, has the form of a target, composed by natural liquid gallium, divided into two zones in such a way, that counting rates in each zone, in this case, the rates of $^{71}$Ge nuclei generation, should be equal in the absence of oscillations. In the presence of oscillations, the rates of gemanium-71 production in each zone will differ from the values, calculated in the non-oscillations scenario, and besides, these rates may differ from each other.

If, when registering neutrino radiation from the Sun, when the flux density is uniform, the counting rate is proportional to the number of absorbing nuclei in the setup and does not depend on the detector geometry, then in calibration experiments, when an artificial neutrino source is located inside the installation, geometry plays a significant role. For the (3+1) model, it is of interest to consider the geometry of experiment that allows observing different data acquisition rates in two zones. It is reasonable to perform the experiment using a solid-state detector; zinc selenide $Zn^{82}Se$ is a promising material for its construction.

---

[*] Semenov_SV@nrcki.ru

## II. NETRINO PATH LENGTH IN THE PRESENCE OF OSCILLATIONS

The counting rate $\Xi$ in a setup with artificial neutrino source is written as:

$$\Xi = I\sigma n L_\nu \quad (1)$$

Here $I$ – is the source intensity, $\sigma$ - is the neutrino absorption cross section, $n$ – is the concentration of absorbing nuclei, $L_\nu$ -is the neutrino path length in the installation, defined by the expression:

$$L_\nu = \int_V \frac{1}{4\pi r^2} dV, \; r - \text{is the distance to source.}$$

To achieve the highest data acquisition rate, the length $L_\nu$ must be maximum for a given mass of the detector working medium. This is done for spherical symmetry - spherical layer ($r_1$, $r_2$), $L_\nu = r_2 - r_1$. Here $r_1$ is the radius of shielding surrounding the source, which can be considered a point source. The BEST experiment uses a symmetric cylinder geometry, h=D [7]. In this case, the intensity of data accumulation is close to the maximum possible, which is specific to spherical shape. This is due to the fact that, for the given mass of absorbing isotope, the path length of the neutrino for a symmetric cylinder and a sphere differs little from each other [8]. In the case of point source in the center

$$\frac{L_{\nu,cil}}{L_{\nu,sph}} = \left(\frac{2}{3}\right)^{1/3} \left[\frac{\pi}{4} + \frac{1}{2} \ln 2\right] = 0.988$$

It should be noted, that calculations based on spherical symmetry allow one to model with a sufficient precision the processes occurring in the setup, which is a symmetrical cylinder.

From the performed calibration experiments it follows that

$$\eta_{GALLEX+SAGE+B} = \frac{\Xi_{ex}}{\Xi_{theor}} = 0.80 \pm 0.05$$

Here $\Xi_{theor}$ – is the calculated counting rate in the non-oscillations scenario.

In the (3+1) model with one sterile neutrino the neutrino flux density at a distance r from the source is

$$\Phi(r) = P(r)\frac{I}{4\pi r^2}, \text{ where}$$

$$P(E_\nu, r) = 1 - \sin^2 2\theta \sin^2\left(1.27\frac{\Delta m^2[eV^2]r[m]}{E_\nu[MeV]}\right)$$

Here $\Delta m^2$ is the squared mass difference between of the mass eigenstates, mixing angle $\theta$ determines the amplitude of the oscillations.

Accordingly, the count rate is written as

$$\Xi = \int_V I\sigma n\, P(r)\frac{1}{4\pi r^2}dV = I\sigma n L_{\nu,osc}$$

$L_{\nu, osc}$ is the neutrino path length in the setup in the presence of oscillations,

$$L_{\nu,osc} = \int_V \frac{P(r)}{4\pi r^2}dV \quad (2)$$

$L_{\nu,osc}$ contains information about the physics of neutrino propagation in the experiment for the oscillations scenario.

### III. GEOMETRY OF EXPERIMENT FOR NEUTRINO SHORT BASE OSCILLATION SEARCHES

Neutrino oscillation length $\delta_{осц}$ equals

$$\delta_{осц}[m] = \frac{\pi E_\nu[MeV]}{1.27\Delta m^2[eV^2]},$$

$$P(E_\nu, r) = 1 - \sin^2 2\theta \sin^2\left(\pi \frac{r}{\delta_{osc}}\right) \quad (3).$$

For a spherical layer of inner radius $r_1$ and thickness $\Delta$, as it follows from (2), (3)

$$L_{\nu,osc}(r_1, \Delta) = \Delta - \sin^2(2\theta) \int_{r_1}^{r_2} \sin^2\left(\pi \frac{r}{\delta_{osc}}\right) dr,$$

$$L_{\nu,osc}(r_1, \Delta) = \Delta\left(1 - \frac{\sin^2(2\theta)}{2}\right) + \sin^2(2\theta) \frac{\delta_{osc}}{4\pi}\left[\sin\left(\frac{2\pi(r_1+\Delta)}{\delta_{osc}}\right) - \sin\left(\frac{2\pi r_1}{\delta_{osc}}\right)\right] \quad (4)$$

Let us consider a two-zone experiment, when a point source inside a shield of radius $r_s$ is surrounded by two concentric layers with equal neutrino path lengths, $L_{\nu,1}=L_{\nu,2}$ [4,5]. For spherical geometry $r_2 - r_1 = r_3 - r_2 = \Delta$, $r_1 \geq r_s$.

In the absence of oscillations, since $L_{\nu,osc,1} = L_{\nu,osc,2} = \Delta$, then $\Xi_1 = \Xi_2$. Convincing proof of the existence of oscillations will be the observation of different count rates in the first and second zones. Thus, for a given value of $\Delta m^2$, it is necessary to find the parameters of the setup, i.e. $r_1$, $\Delta$, so that the ratio

$$\eta = \frac{\Xi_2}{\Xi_1} = \frac{L_{\nu,osc}(r_1 + \Delta, \Delta)}{L_{\nu,osc}(r_1, \Delta)}$$

differs maximally from unity. When calculating $L_{\nu,osc}$, we will use the value $\sin^2(2\theta)=0.42$ [5]. The expression for $\eta$ has the following form:

$$\eta = \frac{\Delta - \sin^2(2\theta) \int_{r_1+\Delta}^{r_1+2\Delta} \sin^2\left(\pi\frac{r}{\delta_{osc}}\right) dr}{\Delta - \sin^2(2\theta) \int_{r_1}^{r_1+\Delta} \sin^2\left(\pi\frac{r}{\delta_{osc}}\right) dr} \quad (5)$$

Therefore, we need to find the extremum of expression (5) in the region $r_1 \subset [r_s, r_s+\delta_{осц}]$, $\Delta \subset [0, \delta_{осц}]$. When determining the search region for the extremum, we take into account that, from the point of view of the economical efficiency of the experiment, $r_1$ and $\Delta$ should have the minimum possible values. Moreover, direct calculation shows that $0.676 \leq \eta \leq 1.472$.

Table 1 shows $r_1$, $\Delta$ and $V_{tot}$ for a number of values of $\Delta m^2$, $2.5 eV^2 \leq \Delta m^2 \leq 20\ eV^2$, consistent with the results of the BEST experiment [5] and cosmological constraints on the mass of light sterile neutrinos [9]. The results correspond to $\eta_{min}=0.676$ or $\eta_{max}=1.472$ depending on which option is complaint with a smaller volume of absorbing material. $V_{tot}$- is the total volume in two zones. Calculations were performed for a $^{51}Cr$ source. $^{51}Cr$ emits neutrinos, 90% of which have an energy of 750 keV. The source is assumed to be a pointlike, surrounded by a shield with an outer radius of $r_s$ = 14 cm.

| $\Delta m^2$, $eV^2$ | $\eta$ | $r_1$, cm | $\Delta$, cm | $V_{tot}$, $m^3$ |
|---|---|---|---|---|
| 2.5 | 1.472 | 27.4 | 27.2 | 2.21 |
| 3 | 1.472 | 22.8 | 22.7 | 1.28 |
| 4 | 1.472 | 17.1 | 17.0 | 0.54 |
| 5 | 0.676 | 33.3 | 13.6 | 0.77 |
| 6 | 0.676 | 27.7 | 11.4 | 0.45 |
| 7 | 0.676 | 23.8 | 9.7 | 0.28 |
| 8 | 0.676 | 20.78 | 8.52 | 0.19 |
| 10 | 0.676 | 16.6 | 6.8 | 0.1 |
| 12 | 1.479 | 21.2 | 5.7 | 0.1 |
| 14 | 1.479 | 18.1 | 4.9 | 0.07 |
| 16 | 1.479 | 15.9 | 4.3 | 0.04 |
| 18 | 1.479 | 14.1 | 3.8 | 0.03 |
| 20 | 0.676 | 17.6 | 3.4 | 0.04 |

Table 1

It should be noted that by changing $r_1$, one can obtain the inverse value of $\eta$, while the counting rates in the zones are reversed. The value of $\Delta$ remains unchanged. For example, for $\Delta m^2=8\ eV^2$, if $r_1=31.7$ cm, then $\eta= 1/1.4792=0.676$. In this case, V=0.35 $m^3$.

## IV. EMPLOYMENT OF $^{82}$Se IN SOLID-STATE NEUTRINO DETECTORS

To construct installations with given values of $r_1$ and $\Delta$, aimed at new types of neutrino searches, it is necessary to develop solid-state neutrino detectors. Various versions of such detectors were considered in [10-13]. The possible presence of background of elastically scattered electrons may complicate their application. Thus, for the $^{51}$Cr source, the maximum kinetic energy of elastically scattered electrons is 550 keV. The kinetic energy of an electron arising from the absorption of a neutrino with an energy $E_\nu$ of 750 keV by the $^{71}$Ga nucleus is $E_\nu - E_{thr}$=520 keV. Therefore, it is advisable to use a target isotope with a lower threshold. Such an isotope is $^{82}$Se. Selenium-82 as an object for solar neutrino registration was considered in [14,15].

In [15], the charge exchange reaction $^{82}$Se($^3$He,t)$^{82}$Br was investigated. Three strong isolated Gamow–Teller transitions to the $1^+$ states of $^{82}$Br* were observed, including a transition to the lowest $1^+$ state, $E_x$ = 75 keV, B(GT) = 0.338(31). The threshold of the reaction

$$\nu_e + {}^{82}\text{Se} \rightarrow e^- + {}^{82}\text{Br}^* (1^+, 75\text{ keV}) \qquad (6)$$

is 170.2 keV. Accordingly, the energy of the resulting electron is 580 keV, i.e. significantly exceeds the energy of background electrons.

The absorption cross sections of neutrinos from artificial sources $^{37}$Ar, $^{51}$Cr, $^{65}$Zn by selenium-82 were calculated in [8]. For $E_\nu$ = 750 keV line, present in $^{51}$Cr neutrino flux, the cross section with the production of $^{82}$Br* with $E_x$ = 75 keV is equal to $(293\pm25)\cdot10^{-46}$ cm$^2$. The error in the cross section for $^{82}$Se is due to the inaccuracy of B(GT) determination, B(GT)=0.338(31). Note that when calculating the count rate according to formula (1) with $^{51}$Cr source, where 90% of neutrinos have the energy 750 keV, the value $\sigma$=264.1$\cdot10^{-46}$ cm$^2$ should be used, which is several times larger than the cross section for $^{71}$Ga, $59\cdot10^{-46}$ cm$^2$.

The large cross-section value and low threshold of the neutrino capture reaction make $^{82}$Se a promising material for conducting calibration experiments. A detector that will register electrons arising in the reaction (6) can be built on the base of scintillating crystals of zinc selenide Zn$^{82}$Se, which are used in experiments for studying double beta decay [16]. Measurements will be carried out in real time.

The main contribution to the background of ZnSe experiment with artificial neutrino source is due to 2ν2β-decay of $^{82}$Se, $T_{1/2}^{2\nu}$=8.6$\cdot10^{19}$ yr [16, 17]. The decay rate is $R_{2\nu2\beta}$=9.6$\cdot10^{-7}$ $Bq/g\ ZnSe$ for 90% enrichment in $^{82}$Se. For 1% resolution of detector background intensity in the region of electron energy $E_e$=580 keV is ~2.9·10$^{-9}$/ $g\ ZnSe$ ·s.

In the process of delayed de-excitation of $^{82}$Br$^*$ two gamma quanta are emitted with energies 29 and 46 keV.

$$^{82}\text{Br*} (1^+, 75 \text{ keV}) \to {}^{82}\text{Br}_{g.s.} + 2\gamma$$

Thus, triple coincidence scheme can be used to reduce $^{82}$Se 2ν2β-background. Let us note, that in the projects, that use LENS technology, based on inverse $^{115}$In beta-decay [18], for solar neutrino registration [18,19] and sterile neutrino searches [20,21], employment of triple coincidence is proposed in order to reject background, due to beta-decay of $^{115}$In, R=0.26$Bq/g\ ZnSe$.

The abundance of $^{82}$Se is 9.19%. Experiments with artificial neutrino sources can be carried out using a detector based on zinc selenide crystals containing selenium enriched in the 82-nd isotope or natural selenium. Enrichment, naturally, leads to a higher counting rate.

In Table 2 counting rates of experiment with Zn$^{82}$Se, 90% enrichment, together with total mass of zinc selenide crystals are presented for a number of $\Delta m^2$ values of interest. The neutrino source is 3.414 MCi $^{51}$Cr. The concentration of $^{82}$Se nuclei is 2.065·10$^{22}$ 1/cm$^3$. The zones dimensions are given in Table 1.

| $\Delta m^2$, eV$^2$ | $\eta = \dfrac{\Xi_2}{\Xi_1}$ | 1-st zone Number of counts per day, $\Xi_1$ | 2-nd zone Number of counts per day, $\Xi_2$ | Mass of Zn$^{82}$Se, tons |
|---|---|---|---|---|
| 2 | 1.472 | 102.6 | 151 | 12.18 |
| 6 | 0.676 | 63.2 | 42.7 | 2.46 |
| 10 | 0.676 | 38 | 25.6 | 0.53 |
| 14 | 1.472 | 18.3 | 27.1 | 0.36 |
| 20 | 0.676 | 19 | 12.8 | 0.21 |

Table 2

Selenium forms a gaseous compound, so the required amount of selenium-82 can be produced in Russia using a cascade of centrifuges [22].

The count rate increases with the growth of source activity. A source emitting higher-energy neutrinos can also be used, resulting in an increased absorption cross section. In [23], it is planned to carry out calibration measurements with $^{71}$Ga, using a $^{58}$Co source. The neutrino flux from $^{58}$Co consists of 98.8% neutrinos with an energy of 1497 keV. The capture cross section for neutrinos with $E_\nu$ = 1497 keV for $^{71}$Ga is 220·10$^{-46}$ cm$^2$. For $^{82}$Se cross section of neutrino absorption, $E_\nu$ = 1497 keV, with excitation of $^{82}$Br* (1$^+$, 75 keV) level equals 821·10$^{-46}$ cm$^2$. For $^{58}$Co - $^{82}$Se experiment maximal kinetic energy of elastically scattered electron equals 1278.7 keV, electron kinetic energy, produced in reaction (6) is $T_e$=1326.8 keV.

The Selena project [13] is developing a setup consisting of modules in which a 5 mm thick layer of amorphous selenium-82 is applied to a matrix made using complementary metal-oxide-semiconductor technology. The detector will be characterized by high spatial resolution, allowing one to determine at what point the electron is generated. $^{51}$Cr source will be applied. The parameters of the zones and, accordingly, $\Delta m^2$ can be determined, based on computer analysis of an array of experimental data, taking into consideration Table 1.

The setup, containing 10 tons of selenium-82 [13] in the case of neutrino source with $r_s$=14 cm for spherical symmetry has the form of spherical layer and, neglecting widths of semiconductor layers, has the thickness 68.5 cm and outer radius 82.5 cm. Thus, according to Table 1, it is possible to search for sterile neutrinos with $\Delta m^2$> 2.5 эВ$^2$. The concentration of $^{82}$Se nuclei for 90% enrichment is 2.92·10$^{22}$ 1/cm$^3$. If chromium-51 source with activity 3.414 MCi [2, 4, 5, 7] is used, then total counting rate in non-oscillation scenario equals 493 electrons per day. To calculate counting rates for Se zones, the results of Table 2 should be multiplied by 1.42. In order to enlarge the range of $\Delta m^2$, which can be investigated, for greater $\Delta m^2$ it is necessary to reduce the width of $^{82}$Se layers.

## V. CONCLUSIONS

For two-zone experiment with $^{51}$Cr source the geometrical parameters have been obtained that ensure the maximum difference in counting rates in zones in the presence of oscillations. Observing this difference would be strong evidence for the existence of sterile neutrinos.
For 3.414 MCi source activity and Zn$^{82}$Se as detector, counting rates in zones are calculated for 2.5eV$^2$≤ $\Delta m^2$≤20 eV$^2$. The region $\Delta m^2$≤2.5eV$^2$ should require greater mass of neutrino absorbing material. To investigate $\Delta m^2$ ≥20 eV$^2$ the source with greater activity and, or source with greater neutrino energy should be applied.

## ACKNOWLEDGMENTS

Author is grateful to A.S. Barabash and Yu.S. Lutostansky for useful discussions

## FUNDING

This work was conducted within a state prescription.

---

1. V.A. Kuzmin, ZHETF 49, 1532 (1965).
2. J.N. Abdurashitov *et al.,* Phys. Rev. C. 80, 015807 (2009).
3. F. Kaether *et al.,* Phys. Lett. B 685, 47 (2010).
4. V.V. Barinov *et al.,* Phys. Rev. Lett. 128, 232501 (2022).
5. V.V. Barinov *et al,* Phys. Rev. C 105, 065502 (2022).
6. S.R. Elliot, V.N. Gavrin, W.C. Haxton, Progr. Part. Nucl. Phys. 134,104082 (2024).
7. V.V. Gorbachev, V.N. Gavrin, and T.V. Ibragimova*,* Phys. Part. Nucl. 49, No 4, 685 (2018).
8. S.V. Semenov, Phys. Atom. Nucl. 85, No 11,1832 (2022).
9. M.A. Acero *et al.,* arXiv 2203.07323 [hep-ex] (2024).
10. V.N. Gavrin, Y.P. Kozlova, E. Veretenkin*,* Nucl. Instr. Meth. A, 466, 119 (2001).
11. K. Zuber*,* Phys. Lett. B, 571, 148 (2003).
12. P. Huber*,* Phys. Rev. D, 107, 096011 (2023).
13. A.E. Chavarria*,* arXiv:2111.00644 [physics.-ins-det] (2021).
14. R.S. Raghavan, Phys. Rev. Lett., 78, 3618 (1997).
15. D. Frekers *et al*., Phys. Rev. C 94, P. 014614 (2016).
16. O. Azzolini et al., Phys. Rev. Lett. 123, 262501 (2019).
17. R. Arnold *et al*, Eur. Phys. J. C, 78, 821 (2018).
18. R.S. Raghavan, Phys. Rev. Lett.37, 259 (1976);
    R.S. Raghavan, 0106054 [hep-ex] (2001).
19. I. R. Barabanov, L.B. Bezrukov, V.I. Gurentsov, G. Ya. Novikova, V.V. Sinev, and E.A. Yanovich, Phys. Atom. Nucl. 86, No. 6, 923 (2023).
20. C. Grieb, J.M. Link, and R.S. Raghavan, Phys. Rev. D 75, 093006 (2007).
21. G. Chauhan, and P. Huber, arXiv: 2507.07397 [hep-ph] (2025).
22. A.V. Tikhomirov, in *Proceedings of the 4th Symposium on Weak and Electromagnetic*

*Interaction in Nuclei, Osaka, Japan* (World Scientific, 1995);

A.V. Tikhomirov, in *Izotopes: Properties, Production, Application*, Ed. by V.Yu. Baranov (Fizmatlit, Moscow, 2005), p. 208 [in Russian].

23. V.N. Gavrin, V.V. Gorbachev, T.V. Ibragimova, V.A. Matveev, arXiv: 2501.08127 [hep-ex] (2025).

+

1111